\documentclass[a4paper]{amsart}
\usepackage[utf8]{inputenc}
\usepackage{amssymb,amsmath,amsthm,eucal,lmodern,mathrsfs,mathtools,graphicx,tikz,array}
\usepackage[T1]{fontenc}
\usepackage[colorlinks = true, linkcolor = teal, citecolor = blue, anchorcolor = blue]{hyperref}
\usepackage[makeroom]{cancel}
\usepackage{dsfont} 
\usepackage[absolute]{textpos}  

  \calclayout

\usepackage{thmtools}

\declaretheorem[name=Remark, style=definition, numbered=no]{rem}
\declaretheorem[name=Theorem, parent=section]{thm}

\newcommand{\mbb}{\mathbb} \newcommand{\mf}{\mathfrak} \newcommand{\mc}{\mathcal}  \newcommand{\on}{\operatorname} 
  \newcommand{\R}{\mathbb{R}}  
\newcommand{\slot}{\;\cdot\;} \newcommand{\w}[1]{\wedge^{\!#1}\,} \renewcommand{\dh}{\hat d\hspace{-.5mm}} \newcommand{\vv}[1]{\vec{#1\mkern-1mu}\mkern 1mu}

\usetikzlibrary{matrix,arrows,decorations.pathmorphing,decorations.markings,calc,patterns}
\tikzset{->-/.style={decoration={
  markings,
  mark=at position #1 with {\arrow{stealth'}}},postaction={decorate}}}
\tikzset{proj_empty/.style={circle,fill=white,inner sep=0pt}}
\tikzset{proj0/.style={circle,fill=white,draw=black,inner sep=0pt}}
\tikzset{proj1/.style={circle,fill=black,draw=black,inner sep=0pt}}
\tikzset{proj2/.style={circle,fill=blue,draw=blue,inner sep=0pt}}
\tikzset{proj3/.style={circle,fill=gray,draw=gray,inner sep=0pt}}

\title{Exceptional algebroids and type IIA superstrings}
\author{Ond\v rej Hul\' ik}
\address{Theoretische Natuurkunde, Vrije Universiteit Brussel, Pleinlaan 2, B-1050 Brussels, Belgium}
\email{ondra.hulik@gmail.com}
\author{Fridrich Valach}
\address{Department of Physics, Imperial College London\\Prince Consort Road, London, SW7 2AZ, UK}
\email{f.valach@imperial.ac.uk}

\begin{document}

\begin{textblock}{5}(14,2.5)
\noindent Imperial/TP/22/FV/1
\end{textblock}

\maketitle
\begin{abstract}
  We study exceptional algebroids in the context of warped compactifications of type IIA string theory down to $n$ dimensions, with $n\le 6$. In contrast to the M-theory and type IIB case, the relevant algebroids are no longer exact, and their locali moduli space is no longer trivial, but has 5 distinct points. This relates to two possible scalar deformations of the IIA theory. The proof of the local classification shows that, in addition to these scalar deformations, one can twist the bracket using a pair of 1-forms, a 2-form, a 3-form, and a 4-form. Furthermore, we use the analysis to translate the classification of Leibniz parallelisable spaces (corresponding to maximally supersymmetric consistent truncations) into a tractable algebraic problem. We finish with a discussion of the Poisson--Lie U-duality and examples given by tori and spheres in 2, 3, and 4 dimensions.
\end{abstract}

\section{Introduction}
Recently, in \cite{BHVW} a new class of geometric objects --- so called \emph{$G$-algebroids} --- was introduced in order to provide a common ground for several types of structures describing the symmetries and dynamics of string and M-theory. Taking $G$ to be the orthogonal group, one recovers the Courant algebroids \cite{LWX}, while the general linear group leads to Lie algebroids \cite{Pradines}. Taking $G$ corresponding to non-compact exceptional Lie groups (instead of $G$-algebroids we then talk simply about \emph{exceptional algebroids} or \emph{elgebroids}), one reproduces in particular the Leibniz algebroids from \cite{PW,Baraglia,CSCW}, used in the study of M-theory and type II string theory compactifications.\footnote{More precisely, these algebroids are used in the study of the corresponding low-energy effective theories (supergravities). However, for simplicity (and in order to match the existing literature) we will simply stick to the terms M-theory/string theory instead.}

The more detailed study of exceptional algebroids was the main focus of the works \cite{BHVW,BHVW2} --- the former focused on the M-theory setup, while the latter was devoted to the type IIB case. In both cases, a local classification result was proved, a method for constructing Leibniz parallelisable spaces \cite{LSCW} was provided, and a general notion of the Poisson--Lie U-duality was studied, extending the exceptional Drinfeld algebra construction of \cite{Sakatani,MT}.

Although quite distinct as far as the technical details are concerned, the M-theory and type IIB cases nonetheless have one thing in common --- they both correspond to a certain ``minimal'' class of exceptional algebroids, called \emph{exact}. The purpose of the present text is to complete the ``triangle'' and provide a detailed derivation of the analogous results in the type IIA case.

In more detail, we describe the relevant subclass of exceptional algebroids, called \emph{type IIA algebroids} and prove that they locally correspond to a type IIA version of the \emph{exceptional tangent bundle} \cite{Hull,CSCW,BCKT,CSCW2,BMP,HS2} (Subsection \ref{subsec:typeIIA} and Theorem \ref{thm:class}). We then study the relation between embedding tensors and Leibniz parallelisations. We show which embedding tensors define such a parallelisation --- they correspond to a pair of an \emph{elgebra} (exceptional algebroid over a point) together with a suitable \emph{coisotropic} subalgebra, satisfying some mild conditions (Theorem \ref{thm:lift}). This result gives a simplification and a slight refinement of the result of Inverso \cite{Inverso}. We then describe the Poisson--Lie U-duality phenomenon and its compatibility with the supergravity equations. We finish by discussing examples of Leibniz parallelisations over tori, $S^2$, $S^3$, and $S^4$, and we explain how the ordinary U-duality and generalised Yang--Baxter deformations fit in our framework.

In addition to being more technically involved, the type IIA setup exhibits an important difference from the M-theory and type IIB case --- the local moduli space of the type IIA exceptional algebroids is not trivial, but consists of 5 points (see the picture in Theorem \ref{thm:class}). This corresponds to two deformations of the type IIA theory \cite{Romans,HLW}. Also, similarly to the IIB case, the ``algebraic calculation'' of the possible twists of the bracket reveals the option of having a non-physical twist by a vector field. This complication results in an extra trace condition in Theorem \ref{thm:lift}, which is not present in the M-theory case \cite{BHVW} (but it appears in the type IIB case \cite{BHVW2}).

\subsection*{Acknowledgement}
The authors would like to thank Alex Arvanitakis, Mark Bugden, Yuho Sakatani, and Daniel Waldram for helpful discussions and comments. O.\,H.\ was supported by the FWO-Vlaanderen through the project G006119N and by the Vrije Universiteit Brussel through the Strategic Research Program ``High-Energy Physics''. F.\,V.\ was supported by the Early Postdoc Mobility grant P2GEP2\underline{\phantom{k}}188247 and the Postdoc Mobility grant P500PT\underline{\phantom{k}}203123 of the Swiss National Science Foundation.

\section{Linear algebra}
\subsection{Exceptional algebras}\label{subsec:excalg}
In order to define exceptional algebroids, we first discuss the neccessary algebraic prerequisites, following \cite{BHVW}. The central role will be played by Lie algebras $E_{n(n)}$, for $n\in\{2,\dots,6\}$, given by the following table:

\begin{center}\begin{tabular}{c|c|c|c|c|c}
$n$ & 6 & 5 & 4 & 3 & 2\\ \hline
$E_{n(n)}$ & $E_{6(6)}$ & $Spin(5,5)$ & $SL(5,\mathbb R)$ & $SL(3,\mathbb R)\times SL(2,\mathbb R)$ & $SL(2,\mathbb R)\times \mathbb R^+$
\end{tabular}\end{center}

Here $E_{6(6)}$ is the connected and simply connected Lie group whose Lie algebra $\mf e_{6(6)}$ is the split real form of the exceptional complex Lie algebra of rank 6. The cases with $n=5$, $4$, and $3$ also correspond to split real forms of complex Lie algebras which, although not exceptional, belong to a certain natural generalisation of the exceptional family, c.f.\ their Dynkin diagrams below.

For each of these groups, there are two representations, labeled $E$ and $N$, which will be of particular interest to us. We list these below the Dynkin diagrams of the (semisimple part of) $E_{n(n)}$. Both $E$ and $N$ correspond to fundamental or trivial representations of (the simple factors of) $E_{n(n)}$. The nodes corresponding to $E$ and $N$ are marked by black and blue colour, respectively.\footnote{Note that a more unified description can be obtained if we decompose everything in terms of a subalgebra $\mf{gl}(n,\mathbb R)\subset \mf{e}_{n(n)}\oplus\R$, as will be shown in Subsection \ref{subsec:Mdecomp}.}

\begin{center}\begin{tabular}{ccccc}
  $\begin{tikzpicture}[scale=0.6, baseline=-0.6ex, thick]
        \draw (0,0) -- (4,0); \draw (2,1) -- (2,0); \node at (2,1) [proj0] {\hphantom{+}}; \node at (0,0) [proj1] {\hphantom{+}}; \node at (4,0) [proj2] {\hphantom{+}};
        \foreach \x in {1,...,3}
            \node at (\x,0) [proj0] {\hphantom{+}};
        \end{tikzpicture}$ &
        $\begin{tikzpicture}[scale=0.6, baseline=-0.6ex, thick]
        \draw (0,0) -- (3,0); \draw (1,1) -- (1,0); \node at (1,1) [proj0] {\hphantom{+}}; \node at (0,0) [proj1] {\hphantom{+}}; \node at (3,0) [proj2] {\hphantom{+}};
        \foreach \x in {1,...,2}
            \node at (\x,0) [proj0] {\hphantom{+}};
        \end{tikzpicture}$ &
        $\begin{tikzpicture}[scale=0.6, baseline=-0.6ex, thick]
        \draw (0,0) -- (2,0); \draw (0,1) -- (0,0); \node at (0,1) [proj0] {\hphantom{+}}; \node at (0,0) [proj1] {\hphantom{+}}; \node at (2,0) [proj2] {\hphantom{+}};
        \foreach \x in {1,...,1}
            \node at (\x,0) [proj0] {\hphantom{+}};
        \end{tikzpicture}$ &
        $\begin{tikzpicture}[scale=0.6, baseline=-0.6ex, thick]
        \draw (1,0) -- (2,0); \node at (0,1) [proj1] {\hphantom{+}}; \node at (1,0) [proj1] {\hphantom{+}}; \node at (2,0) [proj2] {\hphantom{+}};
        \end{tikzpicture}$ &
        $\begin{tikzpicture}[scale=0.6, baseline=-0.6ex, thick]
        \node at (1,0) [proj1] {\hphantom{+}};
        \end{tikzpicture}$
        \\ [.3cm]
        $\mathbf{27}$, $\mathbf{27}'$ & $\mathbf{16}$, $\mathbf{10}$ & $\mathbf{10}$, $\mathbf 5'$ & $(\mathbf 3,\mathbf 2)$, $(\mathbf 3',\mathbf 1)$ & $\mathbf 2_ 1\oplus \mathbf 1_{-2}$, $\mathbf 2_{-1}$
\end{tabular}\end{center}

In fact, we will be interested in the slightly larger group $G:=E_{n(n)}\times \mathbb R^+$. We will consider $E$ and $N$ as $G$-representations, by giving them $\mathbb R^+$-weights 1 and 2, respectively.

Crucially, we have two symmetric $G$-equivariant maps 
\begin{equation}\label{eq:twomaps}
E\otimes E \to N\quad\text{ and }\quad E^*\otimes E^*\to N^*,
\end{equation}
satisfying the following property: First, taking the dual of the second map we get a map $N\to E\otimes E$. Let now $\pi'\colon E^*\otimes E \to E^*\otimes E$ be the partial dual of the composition $E\otimes E\to N\to E\otimes E$. Finally, set $\pi:=1-\pi'$. We then have
\begin{equation}\label{eq:metacond}
  \on{im}\pi\subset \mf g,
\end{equation}
where $\mf g$ is seen as a subalgebra of $E^*\otimes E$.

To avoid complicated notation, we will not give the maps from \eqref{eq:twomaps} specific names, but will instead refer to them using a subscript (signifying a projection). For instance the image of $u\otimes v\in E\otimes E$ under the first map will be denoted simply by $(u\otimes v)_N$. We will use the same type of notation also when dealing with (partial) duals of these maps. For instance, the image of $\xi\otimes n \in E^*\otimes N$ under the map $E^*\otimes N\to E$ (which is a partial dual of $N\to E\otimes E$) will be denoted by $(\xi\otimes n)_E$.

One should think of the maps \eqref{eq:twomaps} as a generalisation of the ordinary inner product, which would correspond to taking the group $O(p,q)$ with $E$ and $N$ being the vector and scalar representations, respectively. This analogy also motivates the following definitions.

We say a subspace $V\subset E$ is 
\begin{itemize}
  \item \emph{isotropic} if $(V\otimes V)_N=0$
  \item \emph{coisotropic} if $(V^\circ\otimes V^\circ)_{N^*}=0$, where $V^\circ\subset E^*$ stands for the annihilator of $V\subset E$.
\end{itemize}
Note that coisotropic spaces can be equivalently characterised by the property $(V^\circ\otimes N)_E\subset V$, see \cite{BHVW}.\footnote{We remark that this definition of isotropy can be seen as a distant relative of the isotropy studied in the context of higher Dirac structures \cite{Zambon}, which is defined for a different pair of $E$, $N$.}

Finally, note that for any $n>2$, we have the relation
\begin{equation}\label{eq:lambda}
  \on{Tr}\pi(A)=\lambda\on{Tr}A,\qquad \forall A\in\on{End}(E),
\end{equation}
where $\lambda=-\tfrac{\dim E}{9-n}$. For clarity and later reference, we give a table of the corresponding values:

\begin{center}\begin{tabular}{c|c|c|c|c}
$n$ & 6 & 5 & 4 & 3 \\ \hline
$-\lambda$ & $9$ & $4$ & $2$ & $1$
\end{tabular}\end{center}

\subsection{M-theoretic decomposition}\label{subsec:Mdecomp}
  In order to be more explicit, let us perform the decomposition under a Lie subalgebra $\mf{gl}(n,\mathbb R)\subset \mf g$. We then obtain
    \begin{align*}
        \mf g&= \R\oplus \mf{gl}(T)\oplus \w{3}T\oplus\w{6}T\oplus\w{3}T^*\oplus\w{6}T^*,\\
        E&=T\oplus \w{2}T^*\oplus\w{5}T^*,\\
        N&=T^*\oplus\w{4}T^*\oplus(T^*\otimes \w{6}T^*),
    \end{align*}
    with $T:=\R^n$.

The first decomposition is arranged such that $\mf{gl}(T)$ acts in the standard way on all the summands, while $\R$ acts with weights 0, 1, 2 on $\mf g$, $E$, $N$, respectively. Denoting the forms and multivectors in the first line by $a$ and $w$ (with subscripts indicating the degree of the form/multivector), and an element of $E$ by $u=X+\sigma_2+\sigma_5$, the remaining parts of the action of $\mf g$ on $E$ are
    \[w_3\cdot u=i_{w_3}(\sigma_2+\sigma_5),\quad w_6\cdot u=-i_{w_6}\sigma_5,\quad a_3\cdot u=i_Xa_3+a_3\wedge \sigma_2,\quad a_6\cdot u=i_Xa_6.\]
    
    The nonzero bits of the symmetric map $E\otimes E\to N$ are given by
    \[(X\otimes\sigma_2)_N=i_X\sigma_2,\quad (X\otimes\sigma_5)_N=i_X\sigma_5,\quad (\sigma_2\otimes\sigma_2')_N=-\sigma_2\wedge\sigma_2',\quad (\sigma_2\otimes\sigma_5)_N=\sigma_2\bar{\otimes}\sigma_5,\]
    where $\sigma_2\bar\otimes \sigma_5\in T^*\otimes \w{6}T^*\cong \on{Hom}(T,\w{6}T^*)$ is defined by $(\sigma_2\bar\otimes \sigma_5)(X)=(i_X\sigma_2)\wedge \sigma_5$.
    
    The map $E^*\otimes E^*\to N^*$ is given by precisely analogous formulas, up to an (unimportant) overall factor which is fixed by the condition \eqref{eq:metacond}, see \cite{BHVW}. In other words, using a suitable inner product on $T$ to identify $E\cong E^*$ and $N\cong N^*$, the two maps \eqref{eq:twomaps} coincide. In particular, there is a bijection\footnote{depending on the choice of the inner product} between the possible isotropic subspaces of dimension $k$ and coisotropic subspaces of codimension $k$.

\subsection{Structure of isotropic subspaces}\label{subsec:isotropics}
  The space of possible isotropic subspaces has an interesting structure, which can be captured by the following Hasse diagram (drawn for all $n\in\{2,\dots,6\}$). This is to be read as follows:
  
  The nodes in the $i$-th line (counted from the bottom) represent isotropic subspaces of dimension $i$, up to an action of an element of $G$. If an isotropic subspace is a subspace of a larger isotropic subspace, the corresponding nodes are linked by an up-going line. Black nodes correspond to maximally isotropic subspaces. Because of the above bijection, these diagrams also capture the structure of coisotropic subspaces.
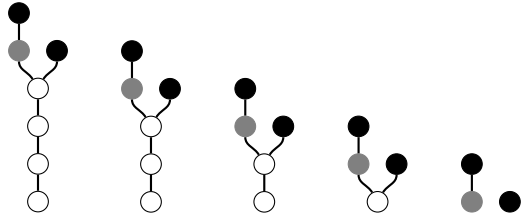
\[\begin{tikzpicture}
  \node (m) at (0,.5) [proj1] {\hphantom{+}}; \node (iib) at (.5,0) [proj1] {\hphantom{+}}; \node (iia) at (0,0) [proj3] {\hphantom{+}};
  \foreach \x in {1,...,4} \node (\x) at (0.25,-\x/2) [proj0] {\hphantom{+}};
  \draw [thick] (iia) to [out=-90, in=120] (1); \draw [thick] (iib) to [out=-90, in=60] (1); \draw [thick] (m) to (iia);
  \draw [thick] (1) to (2); \draw [thick] (2) to (3); \draw [thick] (3) to (4);
\end{tikzpicture}\qquad
\begin{tikzpicture}
  \node (m) at (0,.5) [proj1] {\hphantom{+}}; \node (iib) at (.5,0) [proj1] {\hphantom{+}}; \node (iia) at (0,0) [proj3] {\hphantom{+}};
  \foreach \x in {1,...,3} \node (\x) at (0.25,-\x/2) [proj0] {\hphantom{+}};
  \draw [thick] (iia) to [out=-90, in=120] (1); \draw [thick] (iib) to [out=-90, in=60] (1); \draw [thick] (m) to (iia);
  \draw [thick] (1) to (2); \draw [thick] (2) to (3);
\end{tikzpicture}\qquad
\begin{tikzpicture}
  \node (m) at (0,.5) [proj1] {\hphantom{+}}; \node (iib) at (.5,0) [proj1] {\hphantom{+}}; \node (iia) at (0,0) [proj3] {\hphantom{+}};
  \foreach \x in {1,...,2} \node (\x) at (0.25,-\x/2) [proj0] {\hphantom{+}};
  \draw [thick] (iia) to [out=-90, in=120] (1); \draw [thick] (iib) to [out=-90, in=60] (1); \draw [thick] (m) to (iia);
  \draw [thick] (1) to (2);
\end{tikzpicture}\qquad
\begin{tikzpicture}
  \node (m) at (0,.5) [proj1] {\hphantom{+}}; \node (iib) at (.5,0) [proj1] {\hphantom{+}}; \node (iia) at (0,0) [proj3] {\hphantom{+}};
  \foreach \x in {1,...,1} \node (\x) at (0.25,-\x/2) [proj0] {\hphantom{+}};
  \draw [thick] (iia) to [out=-90, in=120] (1); \draw [thick] (iib) to [out=-90, in=60] (1); \draw [thick] (m) to (iia);
\end{tikzpicture}\qquad
\begin{tikzpicture}
  \node (m) at (0,.5) [proj1] {\hphantom{+}}; \node (iib) at (.5,0) [proj1] {\hphantom{+}}; \node (iia) at (0,0) [proj3] {\hphantom{+}};
  \draw [thick] (m) to (iia);
\end{tikzpicture}\]

  We see that there exists a unique (up to the $G$-action) $n$-dimensional isotropic subspace, i.e.\ $n$-dimensional isotropic subspaces form a single orbit of $G$. In terms of the $M$-theoretic decomposition, this corresponds to $T\subset E$. Correspondingly, coisotropic subspaces of codimension $n$ form a single $G$-orbit — they are all equivalent to $\w{2}T^*\oplus \w{5}T^*\subset E$. Any coisotropic subspace from this orbit will be called \emph{type M}.
  
  In contrast, $n-1$-dimensional isotropic subspaces form two $G$-orbits. One of this is maximal and the other one is not (it can be enlarged to an $n$-dimensional isotropic subspace). This can be most readily seen in the $n=2$ case, where we have $E=T\oplus \w{2}T^*$, with $T=\mathbb R^2$: one-dimensional isotropic subspaces are either given by any 1-dimensional subspace of $T$, or by the 1-dimensional space $\w{2}T^*$ (the former can be enlarged to $T$).
  
  Correspondingly, coisotropic subspaces of codimension $n-1$ are of two types:
  \begin{itemize}
    \item the ones which do not contain any smaller coisotropic subspace, called \emph{type IIB} (depicted by a black node)
    \item those that do contain a smaller coisotropic subspace, called \emph{type IIA} (depicted by a grey node).
  \end{itemize}
  
  Note that one can think of the above Hasse diagrams as capturing the structure of toroidal compactifications of type IIA/B string theory and M-theory, or as capturing the structure of maximally supersymmetric theories in various dimensions. For instance, they show that reducing the type IIA and IIB string theories on a single circle leads to the same theory.
  
  Exceptional algebroids corresponding to the type M and IIB cases were discussed in \cite{BHVW} and \cite{BHVW2}, respectively. In this paper, we concentrate on the third case, given by type IIA.

\subsection{Type IIA subspaces}\label{subsec:iiasubspaces}
  In terms of the M-theoretic decomposition, any type IIA subspace is equivalent (i.e.\ can be related by a $G$-transformation) to the subspace
  \[W=L\oplus \w{2}T^*\oplus\w{5}T^*\subset E,\]
  where $L$ is any 1-dimensional subspace of $E$, spanned by some vector $e\in T$. 

  Furthermore, it will be useful to consider the following Lie algebra
    \[\mf n:=\{\alpha\in\mf g\mid \alpha\cdot E\subset W\}.\]
  A simple calculation shows that
  \begin{equation}\label{eq:n}
    \mf n=\R'\oplus (T^*\otimes L)\oplus\w{3}T^*\oplus\w{6}T^*,\qquad \R':=\{(\tfrac c3,-\tfrac c3\mathds 1)\in\R\oplus\mf{gl}(T)\mid c\in \R\},
  \end{equation}
    with $T^*\otimes L\subset \mf{gl}(T)$. Note that $\R'\subset \mf g$ acts on $T$, $\w{2}T^*$, $\w{5}T^*$ with weights 0, 1, 2, respectively.
  
  Perhaps more explicitly, choosing a decomposition $T=L\oplus\mc T$, with $\mc T\cong \R^{n-1}$, we get a subalgebra $\mf{gl}(\mc T)\subset\mf{gl}(T)$, under which
  \begin{equation}\label{eq:iiadecomp}
  E\cong \mc T\oplus \mc T^*\oplus (\w{0}\mc T^*\oplus\w{2}\mc T^*\oplus\w{4}\mc T^*)\oplus\w{5}\mc T^*,
  \end{equation}
  as follows immediately from the M-theoretic decomposition. The type IIA subspace $W$ then corresponds to
  \begin{equation}\label{eq:Wexplicit}
    W=\mc T^*\oplus (\w{0}\mc T^*\oplus\w{2}\mc T^*\oplus\w{4}\mc T^*)\oplus\w{5}\mc T^*
  \end{equation}
  while its complement $\mc T$ is isotropic.
  
  One can write down a similar decomposition for $N$ and $\mf g$. However, we will not do so --- when doing ``algebraic'' calculations, we will instead work with the M-theoretic decomposition of $E$ (with a chosen subspace $L$).

  Finally, we note an important equivalent characterisation of type IIA subspaces: $W\subset E$ is of type IIA if and only if $\hat W:=(W^\circ\otimes N)_E$ is a subspace of $W$ of codimension 1. This follows from a simple case-to-case check using the classification of coisotropic subspaces. Explicitly, taking the identification \eqref{eq:Wexplicit}, we have
  \[\hat W=\mc T^*\oplus (\w{2}\mc T^*\oplus\w{4}\mc T^*)\oplus\w{5}\mc T^*=\w{2}T^*\oplus\w{5}T^*\subset W,\]
  i.e.\ $\hat W$ misses the $\w{0}\mc T^*$-part.
  \begin{rem}
    The existence of the other inequivalent coisotropic subspace of codimension $n-1$ is reflected in the existence of another inequivalent embedding of $\mf{gl}(\mc T)$ into $\mf g$, under which we get
    \[E\cong \mc T\oplus \mc T^*\oplus (\w{1}\mc T^*\oplus\w{3}\mc T^*\oplus\w{5}\mc T^*)\oplus\w{5}\mc T^*\cong\mc T\oplus (S\otimes \mc T^*)\oplus\w{3}\mc T^*\oplus(S\otimes \w{5}\mc T^*),\]
    with $S:=\R^2$. In this case $(S\otimes \mc T^*)\oplus\w{3}\mc T^*\oplus(S\otimes \w{5}\mc T^*)$ is a type IIB subspace. We will return to this briefly in Theorem \ref{thm:classification}.
  \end{rem}
\section{Exceptional algebroids}
\subsection{Definition}\label{subsec:elgdef}
  We now proceed to the definition of exceptional algebroids (or simply elgebroids) \cite{BHVW}. For simplicity we will use the same letters to denote $G$-representations and the corresponding associated vector bundles. For instance, the crucial object will be a vector bundle $E\to M$, which is assumed to transform in the representation $E$ of the group $G$. Having such a bundle, we then automatically have another vector bundle $N\to M$ (it is associated to the same principal $G$-bundle as $E$).
  
  By definition, an \emph{exceptional algebroid} (or \emph{elgebroid}) is given by the data of a vector bundle $E\to M$ (transforming in the corresponding $G$-representation), together with an $\mathbb R$-bilinear bracket
  \[[\slot,\slot]\colon \Gamma(E)\times\Gamma(E)\to \Gamma(E),\]
  a vector bundle map
  \[\rho\colon E\to TM,\]
  called the \emph{anchor}, and an $\mathbb R$-linear map
  \[\mc D\colon \Gamma(N)\to\Gamma(E),\]
  such that for every $a,b,c\in\Gamma(E)$, $n\in \Gamma(N)$, $f\in C^\infty(M)$ we have
  \begin{align}
    [a,[b,c]]&=[[a,b],c]+[b,[a,c]],\label{eq:jacobi}\\
    [a,fb]&=f[a,b]+(\rho(a)f)b,\label{eq:anchor}\\
    [a,b]+[b,a]&=\mc D(a\otimes b)_N,\label{eq:sym}\\
    \mc D(fn)&=f\mc Dn+(\dh f\otimes n)_E,\label{eq:dfn}\\
    [a,\slot]& \text{ preserves the $G$-structure},\label{eq:pres}
  \end{align}
  where $\dh f:=\rho^T df$. (Here $\rho^T\colon T^*M\to E^*$ is the transpose of $\rho$.)
  
  An elgebroid with $M=*$ (a point) is called an \emph{elgebra}.
  
\begin{rem}
Following an approach from \cite{Severa2} (in the Courant algebroid case), the axiom \eqref{eq:anchor} implies that any $a\in\Gamma(E)$ induces a vector field $s_a$ on the total space $E$ such that
\begin{itemize}
  \item $s_a$ projects down to $\rho(a)$
  \item an infinitesimal $\epsilon$ flow along $s_a$ translates any section $b$ to $b-\epsilon[a,b]$.
\end{itemize}
Axiom \eqref{eq:pres} then says that this vector field, when lifted to the frame bundle of $E$, preserves the subbundle given by the $G$-structure. Equivalently, if $e_\alpha$ is a local $G$-frame of $E$, then so is $e_\alpha':=e_\alpha+\epsilon [a,e_\alpha]$, up to the first order in $\epsilon$ (for any $a$).
\end{rem}

Note that exceptional algebroids form a particular subclass of \emph{vector bundle twisted Courant algebroids} of \cite{GS}.

\subsection{First consequences of the definition}\label{subsec:conseq}
  First, note that since $E\otimes E\to N$ is surjective, $\mc D$ is uniquely determined in terms of the other data (the bundle, G-structure, bracket, and anchor). Furthermore, symmetrising \eqref{eq:jacobi} in $a$ and $b$ we get $[\mc D(a\otimes b)_N,\slot]=0$ and thus
  \begin{equation}\label{eq:central}
    [\mc Dn,\slot]=0,\qquad \forall n\in \Gamma(N).
  \end{equation}
  
  Second, from the axioms \eqref{eq:jacobi} and \eqref{eq:anchor} one easily shows that the anchor intertwines the bracket on $E$ and the commutator of vector fields:
  \begin{equation}\label{eq:intertwine}
    \rho([a,b])=[\rho(a),\rho(b)].
  \end{equation}
  Applying $\rho$ to \eqref{eq:sym} we then get
  \[\rho\circ \mc D=0.\]
  In conjunction with axiom \eqref{eq:dfn} this implies
  \[\rho((\rho^T(T^*M)\otimes N)_E)=0.\]
  In other words, setting
  \[C^0:=T^*M\otimes N,\quad C^1:=E,\quad C^2:=TM,\quad C^3:=0\]
  we get a complex
  \begin{equation}\label{eq:exact}
    C^0\to C^1\to C^2\to C^3.
  \end{equation}
  As shown in \cite{BHVW}, this is equivalent to saying that $\on{ker}\rho\subset E$ is coisotropic at every point.

\subsection{Exact elgebroids}
  We say that the elgebroid is \emph{exact} if \eqref{eq:exact} is an exact sequence, i.e.\ if the homologies $H^1$ and $H^2$ vanish. In this case we have the following classification result \cite{BHVW,BHVW2}.
  \begin{thm}\label{thm:classification}
    Exact elgebroids have locally one of the two following forms. Either
    \[E\cong TM\oplus \w{2}T^*M\oplus\w{5}T^*M,\qquad \dim M=n\]
    \[[X+\sigma_2+\sigma_5,X'+\sigma_2'+\sigma_5']=\mc L_X X'+(\mc L_X \sigma_2'-i_{X'}d\sigma_2)+(\mc L_X \sigma_5'-i_{X'}d\sigma_5-\sigma_2'\wedge d\sigma_2),\]
    or
    \[E\cong \mc TM\oplus (S\otimes \mc T^*M)\oplus \w{3}\mc T^*M\oplus (S\otimes \w{5}\mc T^*M),\qquad \dim M=n-1\]
    \begin{align*}
        [X+\vv\sigma_1+\sigma_3+\vv\sigma_5,X'+\vv\sigma_1'+\sigma_3'+\vv\sigma_5']&=\mc L_X X'+(\mc L_X\vv \sigma_1'-\iota_{X'}d\vv\sigma_1)+(\mc L_X \sigma_3'-\iota_{X'}d\sigma_3+\epsilon_{ij}d\sigma_1^i\wedge {\sigma'_1}^{\!j})\\
        &+(\mc L_X\vv\sigma_5'-\iota_{X'}d\vv\sigma_5+d\sigma_3\wedge \vv\sigma_1'-d\vv\sigma_1\wedge\sigma_3').
    \end{align*}
    In both cases the anchor map is given by the projection onto the first factor. In the second case, the arrow signifies the corresponding tensor is valued in $S=\mathbb R^2$ and we used $\mc TM$ for the tangent bundle over an $n-1$-dimensional base manifold. 
  \end{thm}
  These two cases are the \emph{exceptional tangent bundles for M-theory} and \emph{type IIB string}, respectively \cite{Hull,CSCW,BCKT,CSCW2,BMP,HS2}.
\subsection{Type IIA elgebroids}
  We will be interested here in the third case, corresponding to the type IIA string. This requires that the elgebroid is non-exact, in a certain minimal sense. More precisely, we will say that an elgebroid is of \emph{type IIA} if the following two conditions hold at every point on $M$:
  \begin{itemize}
    \item $\dim H^1=1$
    \item $\dim H^2=0$.
  \end{itemize}
  The second condition says simply that $\rho$ is surjective. Setting $W:=\ker{\rho}$, the first condition is equivalent to saying that 
  $(W^\circ\otimes N)_E\subset W$ is a subspace of codimension 1 at every point. (We use the fact that $W^\circ=\on{im}\rho^T$.) Following the discussion from Subsection \ref{subsec:iiasubspaces} we thus conclude:
  
  \vspace{.2cm}
  \emph{An elgebroid is of type IIA iff $\rho$ is surjective and $\on{ker}\rho$ is of type IIA at every point.}
  \vspace{.2cm}

\subsection{Type IIA exceptional tangent bundle}\label{subsec:typeIIA}
  The prime example of a type IIA elgebroid is the \emph{type IIA exceptional tangent bundle}. This is determined by an $n-1$-dimensional manifold $M$ together with locally constant functions $\beta_0$ and $\varphi$ satisfying $\beta_0\varphi=0$ (if $M$ is connected, we have that either $\beta_0=0$ and $\varphi$ is constant, or the other way round). We will use $\mc TM$ to denote the tangent bundle, in order to distinguish the type II and M-theory cases.
  
  We then take the bundle
  \begin{equation*}
    E=\mc TM\oplus \mc T^*M\oplus (\w{0}\mc T^*M\oplus\w{2}\mc T^*M\oplus\w{4}\mc T^*M)\oplus\w{5}\mc T^*M,
  \end{equation*}
  whose sections will be denoted $x+\tau_1+\tau_0+\tau_2+\tau_4+\tau_5$. The bracket is
  \begin{align*}
    [x+\tau_1+\tau_0+\tau_2+\tau_4+\tau_5&,\;x'+\tau'_1+\tau'_0+\tau'_2+\tau'_4+\tau'_5]=\mc L_x x'+(\mc L_x \tau_1'-i_{x'}d\tau_1)+(\mc L_x \tau_0'-\mc L_{x'}\tau_0)\\
    &+(\mc L_x \tau_2'-i_{x'}d\tau_2+d\tau_0\wedge \tau_1'+\tau_0'd\tau_1)+(\mc L_x \tau_4'-i_{x'}d\tau_4+d\tau_1\wedge\tau_2'+d\tau_2\wedge \tau_1')\\
    &+(\mc L_x \tau_5'+d\tau_0\wedge\tau_4'+\tau_0'd\tau_4-d\tau_2\wedge\tau_2')\\
    &+\beta_0(\tau_0\tau_1'+i_{x'}\tau_2+\tau_0\tau_2'-\tau_0'\tau_2+2\tau_0\tau_4'-\tau_2\wedge\tau_2'+2i_{x'}\tau_5+2\tau_0\tau_5'-2\tau_0'\tau_5)\\
    &+\varphi(i_{x'}\tau_1-\tau_1\wedge \tau_1'-\tau_1\wedge \tau_4')
  \end{align*}
  and the anchor is given by the projection onto the first summand.
  
  The parameter $\varphi$ is the Romans mass \cite{Romans} --- the bracket with $\varphi\neq0$ was studied in \cite{CGI,CFPSCW} (for a precursor work see \cite{HK}). The parameter $\beta_0$ corresponds to the deformation from \cite{HLW}, see also \cite{Inverso}.
  In the special case with $\beta_0=\varphi=0$, the exceptional tangent bundle coincides with the ones from \cite{Hull,CSCW,BCKT,CSCW2,BMP,HS2}.

  Note that, just as the type IIA theory can be obtained by a Kaluza--Klein reduction of M-theory, we can obtain (a class of) type IIA exceptional tangent bundles from certain M-theory ones. For instance, if the base manifold of an M-theory exceptional tangent bundle is a product $M\times S^1$, then restricting to the $S^1$-invariant vectors and forms gives a type IIA tangent bundle over $M$. However, not all type IIA exceptional tangent bundles arise this way --- for instance the above $\beta_0$-deformation cannot be obtained by such a reduction (the other parameter $\varphi$ can be obtained from a twist of the M-theory bracket by a 1-form, see \cite{BHVW}). A similar type of reduction was studied in \cite{Sakatani, BTZ} in the case of Leibniz parallelisations (see Subsection \ref{subsec:embed}) . 
\section{Classification}
  Our first goal is to prove the following theorem
  \begin{thm}\label{thm:class}
    Any type IIA elgebroid is locally of the form of the type IIA exceptional tangent bundle with $\beta_0,\varphi\in \{-1,0,1\}$ constant and $\beta_0\varphi=0$.\footnote{As shown in the proof, the values of $\beta_0$ and $\varphi$ are determined up to a multiple, and hence we can set them to lie in the restricted range $\{-1,0,1\}$.}
  Correspondingly, the local moduli space has 5 points:
  
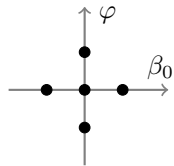
\[\begin{tikzpicture}
  \draw[gray,thick,->] (-1,0) -- (1.1,0); \draw[gray,thick,->] (0,-1) -- (0,1.1); \node at (1,.3) {$\beta_0$}; \node at (.3,1) {$\varphi$};
  \filldraw[black] (0,0) circle (2pt); \filldraw[black] (-.5,0) circle (2pt); \filldraw[black] (.5,0) circle (2pt); \filldraw[black] (0,-.5) circle (2pt); \filldraw[black] (0,.5) circle (2pt);
  \end{tikzpicture}\]
  \end{thm}
  
  
  \subsection{Strategy of the proof}
  We follow the strategy laid out in \cite{BHVW,BHVW2} for the M-theory and type IIB case. In particular, it turns out to be convenient (also for the next Section) to first study a weaker object, called a \emph{type IIA pre-elgebroid}. This is given by the same axioms (including $\dim H^1=1$ and $\dim H^2=0$), except that the Jacobi identity \eqref{eq:jacobi} is replaced by the condition \eqref{eq:intertwine}. We then proceed as follows.
  \begin{enumerate}
    \item We use the axioms to constrain the form of the type IIA pre-elgebroid. This will be locally parametrised by a set of tensors, which will essentially correspond to the twists.
    \item We use the Jacobi identity to derive a set of algebraic and differential (Bianchi) identities for the twists.
    \item During these procedures, some choices will be required. A change in these choices can be interpreted as a gauge transformation for the twists. We use this freedom to locally remove the twists. Some remnants of the scalar twists will remain and will correspond to the parameters $\beta_0$ and $\varphi$.
  \end{enumerate}
  \subsection{Bundle}
    Assume $E$ is a type IIA pre-elgebroid. First, note that we can locally make the identification
    \begin{equation}\label{eq:identification}
      E\cong \mc TM\oplus \mc T^*M\oplus (\w{0}\mc T^*M\oplus\w{2}\mc T^*M\oplus\w{4}\mc T^*M)\oplus\w{5}\mc T^*M,
    \end{equation}
    with the anchor being simply the projection onto the first factor. This follows by the same argument used in \cite{BHVW,BHVW2}, based on the facts that
    \begin{itemize}
      \item The identification \eqref{eq:iiadecomp} provides a decomposition of $E$ into a direct sum of an isotropic subspace $\mc T$ and a type IIA subspace $\mc T^*\oplus (\w{0}\mc T^*\oplus\w{2}\mc T^*\oplus\w{4}\mc T^*)\oplus\w{5}\mc T^*$, with $\mc T=\mathbb R^{n-1}$.
      \item Choosing an isotropic section of the anchor (i.e.\ a vector bundle map $\iota\colon \mc TM\to E$ s.t.\ $\rho\circ\iota=\on{id}$ and $\on{im}\iota$ is isotropic) we get a decomposition of $E$ into a sum of an isotropic and type IIA subspace $E\cong \on{im}\iota\oplus\on{ker}\rho$.
      \item Any two such decompositions (isotropic + type IIA subspace) are related by the action of some element of $G$.
    \end{itemize}
    From now on, we will assume this identification. We can now also use the formulas from Subsection \ref{subsec:Mdecomp} for the maps $E\otimes E\to N$, etc. It remains to fix the form of the bracket $[\slot,\slot]$.
    
    Note that the identification \eqref{eq:identification} is not unique. Two such identifications differ by an anchor-preserving $G$-transformation, i.e.\ by a transformation belonging to the subgroup $N\subset G$, with Lie algebra $\mf n$. This will be interpreted as a gauge transformation.
  \subsection{Bracket}
    Choose local coordinates $x^i$ on $M$. This leads to a trivialisation of $E$,
    \[E\cong M\times (\mc T\oplus \mc T^*\oplus (\w{0}\mc T^*\oplus\w{2}\mc T^*\oplus\w{4}\mc T^*)\oplus\w{5}\mc T^*),\]
    where now we take again $\mc T:=\mathbb R^{n-1}$. In other words, \[\Gamma(E)\cong C^\infty(M)\otimes (\mc T\oplus \mc T^*\oplus (\w{0}\mc T^*\oplus\w{2}\mc T^*\oplus\w{4}\mc T^*)\oplus\w{5}\mc T^*)\]
    and we have a well defined action of vector fields and of the differential $d$ on $\Gamma(E)$ --- they leave the $\mc T\oplus \mc T^*\oplus (\w{0}\mc T^*\oplus\w{2}\mc T^*\oplus\w{4}\mc T^*)\oplus\w{5}\mc T^*$-part intact and only act on $C^\infty(M)$.

    Using this trivialisation, from the axioms \eqref{eq:anchor}--\eqref{eq:dfn} it follows that
    \[[a,b]-\rho(a)b+\pi(\hat d a)b\]
    is $C^\infty(M)$-linear in both $a$ and $b$. In other words, we can write
    \begin{equation}\label{eq:localexpr}
      [a,b]=\rho(a)b-\pi(\hat d a)b+A(a)b,
    \end{equation}
    where $A\colon E\to \on{End}(E)$ is a tensor. Furthermore, the axiom \eqref{eq:pres} ensures that in fact at every point of $M$ we can see $A$ as a map of two vector spaces,
    \[A\colon E\to \mf g.\]
    
    In fact, we can make an even stronger constraint based on the following observation. If $a,b$ are constant sections (w.r.t.\ the above trivialisation) then $\rho(a)$ and $\rho(b)$ are constant vector fields. In particular
    \[\rho(A(a)b)=\rho([a,b])=[\rho(a),\rho(b)]=0,\]
    meaning that $A(a)E\subset \ker \rho$ for any $a$. In other words,
    \begin{equation}\label{eq:aen}
      A\colon E\to \mf n.
    \end{equation}
    
    Similarly, axiom \eqref{eq:dfn} implies that
    \[\mc Dn=(\hat d n)_E+B(n),\]
    where $B\colon N\to E$ is again $C^\infty(M)$-linear.
    Finally, taking $a,b$ constant in \eqref{eq:sym} gives
    \begin{equation}\label{eq:aab}
    A(a)b+A(b)a=B(a\otimes b)_N.
    \end{equation}
    
  \subsection{Algebraic part of the calculation}
    To simplify the problem a bit, we shall use the M-theoretic description and formulas. In other words, we define the bundle $TM:=\mc TM\oplus L$, where $L$ is an auxiliary product line bundle, spanned by a section $e$ (c.f.\ Subsection \ref{subsec:iiasubspaces}). In particular, we have again
    \[E\cong TM\oplus\w{2}T^*M\oplus\w{5}T^*M,\]
    though we have to keep in mind that (despite the notation) $TM$ is now not the tangent bundle of $M$.
    
    A straightforward calculation (see Appendix \ref{sec:appendix}) shows that \eqref{eq:aen} and \eqref{eq:aab} imply
    \begin{align*}
      A(X+\sigma_2+\sigma_5)&=i_X(F_1+F_4)+i_XF_2\otimes e\\
      &+i_{e}i_{\psi} \sigma_2+i_{\psi}\sigma_2\otimes e-(F_1\wedge\sigma_2-F_2\wedge i_{e}\sigma_2)-F_4\wedge \sigma_2\\
      &+i_{\psi} i_{e} \sigma_5-(2F_1\wedge \sigma_5-F_2\wedge i_{e}\sigma_5)
    \end{align*}
    for a set of twists
    \[F_1\in \Gamma(T^*M), \qquad F_2\in \Gamma(\w{2}T^*M), \qquad F_4\in \Gamma(\w{4}T^*M),\qquad \psi\in \Gamma(TM).\]
    This corresponds, in terms of vectors and differential forms on $M$, to a pair of scalars, a pair of 1-forms, a 2-form, a 3-form, a 4-form, and a vector.
    
  \subsection{Explicit form of the bracket}
    In order to be able to use the differential geometric formulas on $TM$ (instead of $\mc TM$), we replace the space $M$ by $M\times \R$, with the coordinate $y$ on $\R$. At every point, the tangent bundle of this space coincides with $TM$, where we identify $\partial_y$ and $e$. For simplicity, we will drop the index $y$ and write just $\partial$.
    
    On the other hand, as the $y$-direction is non-physical, the coefficient functions $X^i$ and $\sigma_{i\dots j}$ will be taken to be independent of $y$. Thus we have for instance $X=x+s_0 \partial$, $\sigma_2=s_2+s_1\wedge dy$, $\sigma_5=s_5+s_4\wedge dy$, where $x+s_1+s_0+s_2+s_4+s_5$ can be seen as a section of \eqref{eq:identification}. Sticking to this notation, the bracket determined in the previous subsections is
    \begin{align}
        [X,\slot]&=\mc L_X+i_X F_1+i_XF_2\otimes\partial+i_XF_4\nonumber\\
        [\sigma_2,\slot]&=-d\sigma_2+i_{\partial}i_{\psi}\sigma_2+i_{\psi}\sigma_2\otimes\partial-(F_1\wedge\sigma_2-F_2\wedge i_{\partial}\sigma_2)-F_4\wedge \sigma_2\label{eq:fullbracket}\\
        [\sigma_5,\slot]&=-d\sigma_5+i_{\psi} i_{\partial} \sigma_5-(2F_1\wedge \sigma_5-F_2\wedge i_{\partial}\sigma_5).\nonumber
    \end{align}
    The terms with no twists arise simply from writing explicitly the first two terms on the RHS of \eqref{eq:localexpr} (e.g.\ using the explicit formulas for $\pi$ from \cite{CSCW}).
    Note that when acting on vectors and forms whose components are independent of $y$, the operation $\mc L_{f\partial}$ acts tensorially (without derivatives) --- at every point we have $\mc L_{f\partial}=-df\otimes\partial\in T^*\otimes L\subset\mf n$.
\subsection{Jacobi and Bianchi}\label{subsec:jac}
    Requiring now the Jacobi identity for the bracket \eqref{eq:fullbracket}, one gets constraints for the twists. These consist of the constraint
    \[\psi=\varphi \partial\]
    for some function $\varphi$, and the requirement that the following expressions vanish
    \[\varphi \,i_{\partial}F_1,\qquad d\varphi-2\varphi(i_{\partial}F_2)-\varphi F_1,\]
    \[dF_1-F_2(i_{\partial} F_1),\qquad dF_2-F_2(i_{\partial}F_2)-\varphi(i_{\partial}F_4),\qquad dF_4+F_4\wedge F_1-F_2\wedge (i_{\partial}F_4).\]
    We will see that the latter set of equations gives rise to the Bianchi identities for the twists.
    
    In fact, these constraints/vanishings follow already from the vanishing of $[\mc D n_1,\slot]$ for $n_1\in T^*\subset N$ (which is a simple consequence of the Jacobi identity, see Subsection \eqref{subsec:conseq}).\footnote{A curious exception is the case $n=5$, where one does not obtain the Bianchi identity for $F_4$ in this way. To derive this Bianchi identity one can instead use the Jacobi identity with $a$, $b$, $c$ being vectors.}
    For instance, taking $n_1=fdy$ and keeping only the $\mbb R'$-part of the expression, we are left with
    \[[\mc D(fdy),\slot]_{\mbb R'\text{-part}}=i_{\psi} df+f(\dots),\]
    where the second term contains no derivatives of $f$.
    Thus its vanishing requires $\psi=\varphi \partial$ for some $\varphi$.
    The rest of the constraints is obtained in a similar fashion.
   
    Taking into account that $\psi=\varphi\partial$, the bracket takes the following simpler form — we shall refer to it as the \emph{twisted bracket}.
    \begin{align}
        [X,\slot]&=\mc L_X+i_X F_1+i_XF_2\otimes\partial+i_XF_4\nonumber\\
        [\sigma_2,\slot]&=-d\sigma_2+\varphi\, i_{\partial}\sigma_2\otimes\partial-(F_1\wedge\sigma_2-F_2\wedge i_{\partial}\sigma_2)-F_4\wedge \sigma_2\label{eq:fulltwisted}\\
        [\sigma_5,\slot]&=-d\sigma_5-(2F_1\wedge \sigma_5-F_2\wedge i_{\partial}\sigma_5).\nonumber
    \end{align}
    An explicit form of this bracket in terms of geometric structures and operations on $M$ can be found, together with the relevant Bianchi identities (see below), in Appendix \ref{ap:fullbkt}.
\subsection{Gauge transformations}
    Infinitesimal gauge transformations are parametrised by an element $\xi\in C^\infty(M)\otimes\mf n$, which we write (using \eqref{eq:n}) as $\xi=\xi_0+\hat\xi+\xi_3+\xi_6$, with $\hat\xi:=\xi_1\otimes\partial$. Assuming the above constrained form of $\psi$, and writing $[\slot,\slot]_{\mc F}$ for the twisted bracket with $\mc F=(\varphi,F_1,F_2,F_4)$, we have
    \[\delta_\xi[\slot,\slot]_\mc F\equiv\xi [\slot,\slot]_{\mc F}-[\xi\slot,\slot]_{\mc F}-[\slot,\xi\slot]_{\mc F}=[\slot,\slot]_{\mc F+\delta_\xi \mc F}-[\slot,\slot]_{\mc F}\]
    from which we can read off the following gauge transformations:\footnote{To determine the expressions, it is enough to examine the transformation properties for $[X,Y]$, for the vector part of $[\sigma_2,X]$, and for $[\sigma_2,X]$ with $i_{\partial}\sigma_2=0$. These will determine transformations of $(F_2,F_4)$, $\varphi$, and $F_1$, respectively.}
    \[\delta_{\xi_0}\varphi=-\xi_0\varphi,\qquad \delta_{\xi_0}F_1=-d\xi_0,\qquad \delta_{\xi_0}F_4=\xi_0 F_4\]
    \[\delta_{\hat \xi}\,\varphi=2\varphi (i_{\partial}\xi_1),\qquad \delta_{\hat \xi}F_1=\hat\xi\cdot F_1,\qquad \delta_{\hat\xi}F_2=-d\xi_1+\hat\xi\cdot F_2+F_2(i_{\partial}\xi_1),\qquad \delta_{\hat\xi}F_4=\hat\xi\cdot F_4\]
    \[\delta_{\xi_3}F_2=\varphi \,i_{\partial}\xi_3,\qquad \delta_{\xi_3}F_4=-d\xi_3-F_1\wedge \xi_3+F_2\wedge i_{\partial}\xi_3.\]
    The remaining non-displayed gauge transformations (including the action of $\xi_6$) are all trivial.
\subsection{Type IIA language}
    Decomposing the twists as
    \[F_1=\alpha_1+dy\wedge \beta_0,\qquad F_2=\alpha_2+dy\wedge \beta_1,\qquad F_4=\alpha_4+dy\wedge \beta_3,\]
    the expressions from the beginning of Subsection \ref{subsec:jac} reduce down to the vanishing of
    \[\varphi\beta_0 \qquad d\varphi-2\varphi\beta_1-\varphi\alpha_1\qquad d\beta_0+\beta_1\beta_0\qquad d\alpha_1-\alpha_2\beta_0\qquad d\beta_1\]
    \[d\alpha_2-\alpha_2\wedge\beta_1-\varphi\beta_3\qquad d\beta_3-\beta_3 \wedge\alpha_1-\alpha_4\beta_0+\beta_1\wedge\beta_3\qquad d\alpha_4+\alpha_4\wedge\alpha_1-\alpha_2\wedge\beta_3.\]
    This provides a set of equations for the twists \[\varphi,\;\beta_0,\;\alpha_1,\;\beta_1,\;\alpha_2,\;\beta_3,\;\alpha_4.\] The first one, $\varphi\beta_0=0$, is purely algebraic, while the rest gives differential Bianchi identities.
    Concerning the gauge transformations, writing
    \[\xi_0=a_0,\qquad \xi_1=a_1+dy\wedge b_0,\qquad \xi_3=a_3+dy\wedge b_2,\]
    we get
    \[\delta_{a_0}\varphi=-a_0\varphi,\qquad \delta_{a_0}\alpha_1=-da_0,\qquad \delta_{a_0}\beta_3=a_0\beta_3,\qquad \delta_{a_0}\alpha_4=a_0\alpha_4,\]
    \[\delta_{b_0}\varphi=2b_0\varphi,\qquad\delta_{b_0}\beta_0=-b_0\beta_0,\qquad\delta_{b_0}\beta_1=db_0,\qquad\delta_{b_0}\alpha_2=b_0\alpha_2,\qquad\delta_{b_0}\beta_3=-b_0\beta_3,\]
    \[\delta_{a_1}\alpha_1=-a_1\beta_0,\qquad\delta_{a_1}\alpha_2=-da_1-a_1\wedge\beta_1,\qquad\delta_{a_1}\alpha_4=-a_1\wedge\beta_3,\]
    \[\delta_{b_2}\alpha_2=b_2\varphi,\qquad\delta_{b_2}\beta_3=db_2+b_2\wedge\alpha_1+b_2\wedge\beta_1,\qquad\delta_{b_2}\alpha_4=b_2\wedge\alpha_2,\]
    \[\delta_{a_3}\beta_3=-a_3\beta_0,\qquad\delta_{a_3}\alpha_4=-da_3+a_3\wedge\alpha_1.\]
    
\begin{rem}
    The non-zero-form twists can be associated with field strengths of the bosonic fields appearing in the restriction of the type IIA supergravity to $n-1$ dimensions, with a warp factor. This consists of
    \begin{itemize}
      \item two scalars --- the dilaton and the warp factor
      \item a 1-form and a 3-form (Ramond--Ramond fields)
      \item a 2-form (Kalb--Ramond field).
    \end{itemize}
\end{rem}

\subsection{Removing the twists}
    We now use the gauge transformations to remove most of the twists one by one, taking account of the relevant Bianchi identities.
    The equation $\varphi\beta_0=0$ implies that we can (locally) distinguish two cases --- either $\varphi=0$ or $\beta_0=0$.
    
    Suppose first that $\varphi=0$. We notice that $\beta_1$ is closed. We can thus locally use a (finite version) of the $b_0$-transformation to set $\beta_1$ to zero. We will write this step as $b_0\leadsto\cancel{\beta_1}$. The Bianchi identities then reduce to the vanishing of
    \[d\beta_0\qquad d\alpha_1-\alpha_2\beta_0,\qquad d\alpha_2\qquad d\beta_3-\beta_3 \alpha_1-\alpha_4\beta_0\qquad d\alpha_4+\alpha_4\alpha_1-\alpha_2\beta_3,\]
    while the remaining gauge transformations take the form
    \[\qquad \delta_{a_0}\alpha_1=-da_0,\qquad \delta_{a_0}\beta_3=a_0\beta_3,\qquad \delta_{a_0}\alpha_4=a_0\alpha_4,\]
    \[\delta_{a_1}\alpha_1=-a_1\beta_0,\qquad\delta_{a_1}\alpha_2=-da_1,\qquad\delta_{a_1}\alpha_4=-a_1\wedge\beta_3,\]
    \[\delta_{b_2}\beta_3=db_2+b_2\wedge\alpha_1,\qquad\delta_{b_2}\alpha_4=b_2\wedge\alpha_2,\]
    \[\delta_{a_3}\beta_3=-a_3\beta_0,\qquad\delta_{a_3}\alpha_4=-da_3+a_3\wedge\alpha_1.\]
    Since the Bianchi identity for $\alpha_2$ was now simplified to $d\alpha_2=0$, we can use an appropriate $a_1$-transformation to set $\alpha_2=0$, i.e.\ $a_1\leadsto \cancel{\alpha_2}$. Crucially, note that the $a_1$-transformation will not generate a nonzero value of the twists which were already set to zero (in our case $\beta_1$ and $\varphi$). Continuing this procedure, we can now perform
    \[a_0\leadsto\cancel{\alpha_1},\qquad a_3\leadsto\cancel{\alpha_4}, \qquad b_2\leadsto\cancel{\beta_3}\]
    and we are left with only one non-zero twist $\beta_0$, which is (locally) constant. We now note that we can still scale $\beta_0$ using a constant $b_0$-transformation, and so we can set it to one of the three values: $-1$, $0$, $1$.
    
    Suppose now that $\beta_0=0$. Similarly to above, after performing
    \[b_0\leadsto\cancel{\beta_1},\qquad a_0\leadsto\cancel{\alpha_1},\qquad b_2\leadsto\cancel{\beta_3},\qquad a_3\leadsto\cancel{\alpha_4},\qquad a_1\leadsto\cancel{\alpha_2},\]
    we end up with a single locally constant twist $\varphi$, which can again be scaled via a $b_0$-transformation to one of the values $-1$, $0$, $1$.
    
    Writing the bracket in terms of operations on the manifold $M$ (instead of $M\times\mathbb R$), we obtain precisely the bracket of the type IIA exceptional tangent bundle. This concludes the proof of the Theorem. 

\section{Embedding tensors and Leibniz parallelisations}

  Out second main goal is to study the relation between embedding tensors and Leibniz parallelisations. Again, we follow here the general approach from \cite{BHVW,BHVW2}, where the M-theory and type IIB cases were discussed. 
  \subsection{Definitions}\label{subsec:embed}
  A type IIA elgebroid $E'$ over $M'$ is said to be \emph{Leibniz parallelisable} if there exists a global $G$-frame $e_\alpha$ such that $[e_\alpha,e_\beta]=c_{\alpha\beta}^\gamma e_\gamma$, with constant structure coefficients $c_{\alpha\beta}^\gamma$. Such a choice of frame provides a \emph{Leibniz parallelisation} of the elgebroid. We shall restrict our attention to manifolds $M'$ which are compact and connected.
  
    Given such a parallelisation, we have in particular that $E'$ is a product bundle, $E'\cong M'\times E$, where the vector space $E$ lies in the corresponding representation of $G$. Elements of $E$ correspond to constant sections of $E'$. Since the structure coefficients are constant, $E$ inherits a well-defined bracket. Understanding $E$ as a vector bundle over a point, this defines an exceptional algebroid structure on $E$ (with $\rho=0$), i.e.\ $E$ is an elgebra (see Subsection \ref{subsec:elgdef}).
  
    It was shown in \cite{LSCW} that Leibniz parallelisations correspond to consistent truncations to theories with maximal supersymmetry. The associated elgebras then correspond to the embedding tensors of these lower-dimensional theories.
    
    We now answer the following landscape-type question:
  
    \vspace{.3cm}
    \emph{Which elgebras (embedding tensors) correspond to Leibniz parallelisations?}
  \subsection{Actions of elgebras}
    Assume we have a Leibniz parallelisation over a (compact connected) manifold $M'$. First, note that (as a consequence of the axioms) it can be reconstructed from the following data:
    \begin{itemize}
      \item the base manifold $M'$
      \item the elgebra $E$
      \item the anchor map $\rho'\colon M'\times E\to \mc TM'$.
    \end{itemize}
    Since $\rho'$ is a vector bundle map, we can understand it as a map $\chi\colon E\to \mf X(M'):=\Gamma(\mc TM')$ by taking $\chi(x)_m=\rho((m,x))$ for any $m\in M'$. Because of \eqref{eq:intertwine}, $\chi$ is a map of algebras.   
    In analogy with the case of ordinary Lie algebras, we will thus say that $\chi$ is an \emph{action} of $E$ on $M'$. We conclude that any Leibniz parallelisation gives rise to an action of an elgebra on a manifold $M'$. Conversely, this action determines the parallelisation.
    
    \subsection{From elgebras to Lie algebras}
      Recall that a general elgebra $E$ is not a Lie algebra, since the bracket is not necessarily skew-symmetric. However, the subspace
      \[I:=\on{im}\mc D=\{[a,b]+[b,a]\mid a,b\in E\}\subset E\]
      is a (both-sided) ideal. This follows from \eqref{eq:central} and the fact that
      \[[a,[b,c]+[c,b]]=[[a,b],c]+[b,[a,c]]+[[a,c],b]+[c,[a,b]]=([[a,b],c]+[c,[a,b]])+([b,[a,c]]+[[a,c],b]).\]
      Consequently, the quotient $\mf g_E:=E/I$ inherits a skew-symmetric bracket (satisfying Jacobi identity), i.e.\ it is a Lie algebra.
      
      Note that the projection \[p\colon E\to \mf g_E\] gives a bijection between Lie subalgebras of $\mf g_E$ and subalgebras\footnote{Here we simply mean a linear subspace of $V\subset E$ such that $[V,V]\subset V$.} of $E$ containing $I$. The Lie subalgebra of $\mf g_E$ corresponding to a subalgebra $V\subset E$ will be denoted by $\mf g_V$.\footnote{We reserve the fraktur notation for Lie (sub)algebras.}
      
      Finally, any action $\chi$ of an elgebra $E$ on a manifold induces a corresponding action of $\mf g_E$, since
      \[\chi([a,b]+[b,a])=[\chi(a),\chi(b)]+[\chi(b),\chi(a)]=0.\]
      If the action $\chi$ was transitive ($\chi$ is surjective at every tangent space on $M'$), so is the action of $\mf g_E$.
      
    \subsection{From manifolds to algebras}
      On an exact elgebroid the anchor map is surjective. Thus any Leibniz parallelisation induces a transitive action of a Lie algebra $\mf g_E$ on $M'$. Since we assume $M'$ to be compact and connected, this in turn implies that $M'\cong G_E/H$, where $G_E$ is the connected and simply-connected Lie group integrating $\mf g_E$ and $H\subset G_E$ is a Lie subgroup with a subalgebra $\mf h\subset \mf g_E$.
      
      Using the correspondence between subalgebras of $\mf g_E$ and $E$, we have $\mf h=\mf g_V$ for some subalgebra $V\subset E$ containing $I$. Thus a Leibniz parallelisation gives rise to a pair $(E,V)$, with $I\subset V$. Furthermore, since $\on{ker}\rho'$ is, at the coset of $1\in G_E$, identified with $V$, we see that $V$ has to be a type IIA subspace of $E$.
      
      Conversely, we can ask:
      
      \vspace{.3cm}
      \emph{Which pairs $(E,V)$ of an elgebra and its subalgebra, with $V$ being a type IIA subspace containing $\on{im}\mc D$, correspond to a Leibniz parallelisation?}
      \vspace{.3cm}
      
      This is answered by the following Theorem. A similar result was derived, using different methods, in \cite{Inverso}.
      \begin{thm}\label{thm:lift}
        Let $n>2$. Suppose $E$ is an elgebra and $V\subset E$ a subalgebra such that $\on{im}\mc D\subset V$ and $V$ is a type IIA subspace, with $G_V\subset G_E$ a closed subgroup. Then the pair $(E,V)$ defines a Leibniz parallelisation (over $G_E/G_V$) if and only if the following condition holds:
        \[\on{Tr}_E\on{ad}_a=\tfrac{\lambda}{\lambda-1}\on{Tr}_V\on{ad}_a\qquad \forall a\in \hat V,\]
        where $\hat V:=(V^\circ\otimes N)_E\subset V\subset E$ and $\lambda$ is the constant defined in Subsection \ref{subsec:excalg}.
      \end{thm}
      \begin{rem}
        We assume here that $G_E$ is the connected and simply connected group with the Lie algebra $\mf g_E$. Note that we have, in general, some freedom in the choice of the subgroup $G_V\subset G_E$ corresponding to a given $\mf g_V$. These differ in the number of connected components. For instance, for $\mf g_V=0$ and $G_E=SU(2)$ we can take $G_V=\{1\}$ or $G_V=\{\pm 1\}$, leading to the quotients $SU(2)\cong S^3$ or $SO(3)\cong \mbb{RP}^3$, respectively.
      \end{rem}
  \subsection{Proof of the Theorem}
    Suppose $E$ is an elgebra and $V\subset E$ is a subalgebra which contains $\on{im}\mc D$ and is of type IIA. Supposing the corresponding type IIA elgebroid exists, it necessarily has the form \[E'=E\times M',\qquad M':=G_E/G_V,\] with the anchor and bracket given by the following formulae.
    First,
    \begin{equation}\label{eq:upliftanch}
      \rho'(a)_{[g]}=\left.\tfrac{d}{dt}\right|_{t=0}[e^{-tp(a)}g],
    \end{equation}
    where $[g]\in G_E/G_V$ denotes the coset of $g\in G_E$. (In other words, the anchor is the infinitesimal action of $G_E$ on $G_E/G_V$.) The bracket is
    \begin{equation}\label{eq:upliftbkt}
      [a,b]'=[a,b]+\rho'(a)b-\pi(\hat d a)b.
    \end{equation}
    where $[\slot,\slot]$ is the bracket on $E$ and we have used the identification $\Gamma(E')\cong C^{\infty}(M')\otimes E$.
    
    A simple check reveals that this object always satisfies the axioms of a type IIA pre-elgebroid. It remains to check under what conditions the Jacobi identity is satisfied --- in other words, when the \emph{Jacobiator}
    \[J(a,b,c):=[a,[b,c]']'-[[a,b]',c]'-[b,[a,c]']'\]
    vanishes. We first note that the Jacobiator of constant sections does vanish, as a consequence of the Jacobi identity of $E$. Consequently, the Jacobiator vanishes completely iff it is a tensor (i.e.\ $C^\infty(M')$-linear in all slots).
    
    We know that any type IIA pre-elgebroid has locally the form \eqref{eq:fullbracket}, for some general $\psi$, $F_1$, $F_2$, $F_4$. (For a general pre-elgebroid the twists are not constrained by the Bianchi identities.) A straightforward calculation shows that for such bracket the Jacobiator is a tensor iff $\psi=\varphi\partial$ for some function $\varphi$.
    
    Let now $\hat K:=((\on{ker}\rho')^\circ\otimes N')_{E'}$ be the codimension-one subspace of $\on{ker}\rho'$ (see Subsection \ref{subsec:iiasubspaces}). This corresponds to the subspace
    \[\w{2}T^*M\oplus\w{5}T^*M\cong \mc T^*M\oplus (\w{2}\mc T^*M\oplus\w{4}\mc T^*M)\oplus\w{5}\mc T^*M.\]
    It is now easy to see that the condition $\psi=\varphi\partial$ (for some $\varphi$) corresponds precisely to the following condition:
    \[\on{Tr}_E[a,\slot]'=0,\qquad \forall a\in\Gamma(\hat K).\]
    Note that since $\hat K\subset \on{ker}\rho'$, for $a\in \Gamma(\hat K)$ we have $[a,fb]'=f[a,b]'$ and so $[a,\slot]'$ has a meaningful (pointwise) trace. Using \eqref{eq:upliftbkt} and \eqref{eq:lambda}, the above trace condition is equivalent to
    \[\on{Tr}_E [a_{[g]},\slot]=\lambda\on{Tr}_E(\hat da)_{[g]},\qquad \forall a\in\Gamma(\hat K),\, \forall g\in G_E.\]
    Finally, using \eqref{eq:upliftanch} and relating the expression to the corresponding expression at $g=1$, this is equivalent to (see \cite{BHVW2} for a detailed derivation)
    \[\on{Tr}_E\on{ad}_a=\lambda\on{Tr}_{E/V}\on{ad}_a,\qquad\forall a\in \hat V=(V^\circ\otimes N)_E.\]
    Using $\on{Tr}_{E/V}\on{ad}_a=\on{Tr}_E\on{ad}_a-\on{Tr}_V\on{ad}_a$, we obtain the condition from the statement of the Theorem, which concludes the proof.

\section{Poisson--Lie U-duality and examples}
  Let us now discuss the Poisson--Lie U-duality phenomenon and provide some examples. Some details can be found also in \cite{CGLP}.
  \subsection{Poisson--Lie U-duality}
    Assume we have an elgebra $E$ which admits multiple type IIA subalgebras $V$, which contain $\on{im}\mc D$ and satisfy the trace condition. Then the corresponding Leibniz parallelisable spaces are said to be \emph{Poisson--Lie U-dual}. Poisson--Lie U-duality is a phenomenon introduced in the works \cite{Sakatani,MT} and was given an interpretation in terms of elgebroids in \cite{BHVW}. It generalises both U-duality and Poisson--Lie T-duality of Klim\v c\'ik and \v Severa \cite{KS}. To see one of the implications, let us briefly comment on an aspect of the corresponding supergravity.
    
    Taking a warped compactification of the type IIA supergravity down to $n$-dimensions, we obtain a supergravity which can be succinctly described via a type IIA elgebroid. Ignoring for now the global issues, this essentially corresponds to the exceptional tangent bundle of type IIA. The bosonic degrees of freedom correspond to the reduction of the structure group from $G$ to its maximal compact subgroup $K(G)$ and the dynamics is encoded in the vanishing of a suitable curvature tensor, constructed using (generalised) torsion-free $K(G)$-compatible connections,\footnote{Strictly speaking, instead of passing from $G$ to $K(G)$ one should go to the double cover of $K(G)$ --- this is because in the construction of the curvature tensor we use exceptional analogues of the spinor representations.} see \cite{CSCW} and the references within.
    
    The formulation of the theory in terms of elgebroids immediately implies the compatibility of the supergravity equations of motion with the Poisson--Lie U-duality. This was shown in \cite{BHVW} in the M-theory case --- however, the same result applies also to the type IIA (and type IIB) case. This is because in the type IIA (and also IIB) case one uses exactly the same expressions for the curvature tensors and the same reduction of the structure group as for the M-theory case, see \cite{CSCW} (the only change is that in type II we decompose the representations under the $\mf{gl}(\mc T)$-subalgebra of $\mf g$, instead of $\mf{gl}(T)\subset \mf g$).

  \subsection{Torus}
    We start with the simplest example, taking the elgebra with \[[\slot,\slot]=0.\]
    In this particular example, let us abandon the requirement that $G_E$ is simply connected and take it to be the torus $T^{\dim E}$. In that case, if we take any type IIA subspace $V\subset E$, which corresponds to a closed subgroup of $G_E$, we obtain a Leibniz parallelisation on the quotient $T^n$. Different choices of the type IIA subspace are linked by the action of the exceptional group --- they correspond to different dual tori. This is the standard U-duality.
  
  \subsection{2-sphere}
    Take $n=3$. A particularly nice example of an elgebra corresponds to a Leibniz parallelisation of the 2-sphere, discussed in \cite{CLP}. It can be elegantly obtained from the M-theoretic parallelisation of the group $SU(2)$, using the frame of left-invariant tensors. This has the following form:
    \[E\cong \mf{su}(2)\oplus \w{2}\mf{su}(2)^*,\qquad[X+\sigma_2,X'+\sigma_2']=\on{ad}_X X'+(\on{ad}_X \sigma_2'-i_{X'}\delta\sigma_2),\]
    where $\on{ad}$ denotes the (co)adjoint action on $\mf{su}(2)$ and $\w{2}\mf{su}(2)^*$, and $\delta$ is the Chevalley--Eilenberg differential.
    The subspace $\w{2}\mf{su}(2)^*$ is of type M and it coincides with the ideal $I$. Consequently, taking a $\mf u(1)$-subalgebra of $\mf{su}(2)$, the subspace
    \[V=\mf{u}(1)\oplus \w{2}\mf{su}(2)^*\]
    is of type IIA and contains $I$. As one easily sees, it is also a subalgebra and satisfies the trace condition. Consequently, we get a Leibniz parallelisation on the manifold
    \[G_E/G_V=SU(2)/U(1)\cong S^2.\]
    
    
  \subsection{3-sphere}
    This corresponds to $n=4$. Since the 3-sphere can be identified with the group $SU(2)$, it automatically admits a Leibniz parallelisation. The elgebra is
    \begin{equation*}
    E=\mf{su}(2)\oplus \mf{su}(2)^*\oplus (\w{0}\mf{su}(2)^*\oplus\w{2}\mf{su}(2)^*),
    \end{equation*}
    \begin{align*}
      [x+\tau_1+\tau_0+\tau_2&,\;x'+\tau'_1+\tau'_0+\tau'_2]=\on{ad}_x x'+(\on{ad}_x \tau_1'-i_{x'}\delta\tau_1)+(\on{ad}_x \tau_2'-i_{x'}\delta\tau_2+\tau_0'\delta\tau_1)\\
      &+\beta_0(\tau_0\tau_1'+i_{x'}\tau_2+\tau_0\tau_2'-\tau_0'\tau_2)+\varphi(i_{x'}\tau_1-\tau_1\wedge \tau_1')
    \end{align*}
    for any $\beta_0,\varphi\in\mbb R$. The corresponding type IIA subalgebra is of the form
    \[V=\mf{su}(2)^*\oplus (\w{0}\mf{su}(2)^*\oplus\w{2}\mf{su}(2)^*).\]
    Again, taking different suitable $V$'s results in Poisson--Lie U-dual setups. Requiring that the dual $V$ is still transverse to $\mf{su}(2)\subset E$, the condition $[V,V]\subset V$ can be identified with a generalisation of the Yang--Baxter equation \cite{BDMCS,BGM,Sakatani,MT,MST,BHVW2}.\footnote{This example extends to the more general case, where we replace the Lie algebra $\mf{su}(2)$ by an arbitrary Lie algebra of dimension $n\in\{3,\dots,6\}$. Nevetheless, $\mf{su}(2)$ stands out as the only compact simple Lie algebra in this dimension range.}
  
  \subsection{4-sphere} This is an example from \cite{NVvN}. It corresponds to the $n=5$ case, where we have $\mf g\cong \mf{so}(5,5)\oplus\mbb R$. Under the diagonal embedding of $\mf{so}(5)$ into the maximal compact subalgebra $\mf{so}(5)\oplus\mf{so}(5)\subset \mf g$, the representation $E$ decomposes as $\mathbf{10}\oplus\mathbf 5\oplus\mathbf 1$. (Similarly, $N$ decomposes as $\mathbf 5\oplus\mathbf 5$.) Guided by this decomposition, we define an elgebra
    \[E=\mf{so}(5)\oplus V_5\oplus V_1,\]
    where $V_5:=\mbb R^5$ and $V_1:=\mbb R^1$ correspond to $\mathbf 5$ and $\mathbf 1$, respectively. The bracket is defined as follows:
    \begin{itemize}
      \item the bracket $[a,\slot]$ with an element $a\in\mf{so}(5)$ is given by the $\mf{so}(5)$ representation on $\mathbf{10}\oplus\mathbf 5\oplus\mathbf 1$
      \item for any $a\in V_5\oplus V_1$, $[a,\slot]=0$.
    \end{itemize}
    In particular, we have $\on{im}\mc D=V_5\subset E$. Any subalgebra $V\subset E$ of codimension 4, which contains $\on{im}\mc D$, is necessarily of the form
    \[V=\mf{so}(4)\oplus V_5\oplus V_1,\]
    for some subalgebra $\mf{so}(4)\subset \mf{so}(5)$. It is easy to see that this is coisotropic:
    
    First, identify $\mathbf{10}\cong \w{2}\mbb R^5$, $\mathbf{5}\cong \w{4}\mbb R^5$, and $\mathbf{1}\cong \w{0}\mbb R^5$. Taking the standard inner product on $\mbb R^5$, we get an identification $E^*\cong E$. Then the map $E^*\otimes E^*\to N^*$, restricted to $\mathbf{10}$, becomes
    \[\w{2}\mbb R^5 \otimes \w{2}\mbb R^5 \to \w{4}\mbb R^5,\qquad \omega\otimes\omega'\mapsto \omega\wedge \omega'.\]
    For the above $V$, we have $V^\circ =e\wedge \mbb R^5\subset \w{2}\mbb R^5$, for some vector $e\in\mbb R^5$. This clearly satisfies the coisotropy condition.
    
    Furthermore, consider the subspace $\mf{so}(4)\oplus V_5\subset V$. Its annihilator is identified with $V^\circ\oplus \mathbf 1$. Since any map from $\mathbf 1\otimes \mathbf{10}$ or $\mathbf 1\otimes \mathbf{1}$ into $N=\mathbf 5\oplus\mathbf 5$ is automatically zero, we get \[((V^\circ\oplus \mathbf 1)\otimes (V^\circ\oplus \mathbf 1))_{N^*}=0\]
    and so $\mf{so}(4)\oplus V_5$ is coisotropic (and hence a type M subspace). Thus $V$ is a type IIA subspace.
    
    It also satisfies the trace condition and hence leads to a Leibniz parallelisation of the space
    \[G_E/G_V=(SO(5)\times \mbb R)/(SO(4)\times \mbb R)\cong S^4.\]
    

\section{Conclusions}
  We have shown how the Leibniz algebroids appearing in the study of type IIA string theory fit naturally in the framework of exceptional algebroids. We have proved a classification result which shows that type IIA exceptional bundles admit an abstract characterisation (in terms of a natural asociated sequence). This in especially convenient for the purposes of Poisson--Lie U-duality --- in particular we were lead to an elegant and concise description of this duality as well as to the proof of its compatibility with the supergravity equations of motion.
  
  We have also simplified and refined slightly the result of Inverso \cite{Inverso} concerning the correspondence between embedding tensors and Leibniz parallelisations. This leads to an efficient algorithm for searching for maximally supersymmetric consistent truncations. Although we do not attempt to provide a complete classification of such truncations here, we showed how the known sphere examples from the literature fit into our framework.
  
  From a more mathematical perspective, this article might prove to be of interest due to the fact that --- even though they provide a natural extension of exact Courant algebroids --- the type IIA exceptional algebroids have a nontrivial (local) moduli space, corresponding to two possible deformations of the type IIA theory.
  
  Similarly to \cite{BHVW,BHVW2}, the present work focused solely on maximally supersymmetric consistent truncations. However, the authors believe that similar technology can be applied also to the non-maximally supersymmetric case, where relatively little is known and more systematic theory can help to find new consistent truncations. These investigations are left for a future work.

{\appendix
\section{The calculation}\label{sec:appendix}
Let us write $A=A_0+ (A_1\otimes e)+ A_3+ A_6$ for a map $A\colon E\to \mf n$. We now need to solve the equation \[A(u)v+A(v)u=B(u\otimes v)_N,\] for some map $B=B^1+B_2+B_5\colon N\to E$.
    
    \subsection*{Vectors}
    First, taking $u=X$, $v=Y$ this becomes \[0=A(X)\cdot Y+A(Y)\cdot X=i_Y(A_3(X)+A_6(X))+i_X(A_3(Y)+A_6(Y))+e(i_Y A_1(X)+i_X A_1(Y)),\]
    which implies
    \[A(X)=i_X(F_1+F_4)+i_XF_2\otimes e,\]
    for some $F_1\in T^*$, $F_2\in \w{2}T^*$, $F_4\in \w{4}T^*$.
    
    \subsection*{2-forms}
    Next, taking $u=X$, $v=\sigma_2$, we have
    \[B(i_X\sigma_2)=i_X(F_1+F_4)\wedge\sigma_2-i_XF_2\wedge i_{e}\sigma_2+i_X(A_3(\sigma_2)+A_6(\sigma_2))+e(i_X A_1(\sigma_2)).\]
    For $\sigma_2$ decomposable, there exists $n-2$ independent vectors $X$ which give zero when contracted with $\sigma_2$ and thus (using the above equality) also when contracted with
    \[(F_1+F_4)\wedge\sigma_2-F_2\wedge i_{e}\sigma_2+A_3(\sigma_2)+A_6(\sigma_2).\] But this expression has form degree higher than 2, and hence it has to vanish. We get
    \[A_3(\sigma_2)=-F_1\wedge \sigma_2+F_2\wedge i_{e}\sigma_2,\qquad A_6(\sigma_2)=-F_4\wedge\sigma_2.\]
    Due to the fact that decomposable 2-forms span the whole $\w{2}T^*$, this result is valid for an arbitrary 2-form $\sigma_2$. 
    
    We also see that $B^1(T^*)\subset L$. Writing $B^1(\sigma_1)=-e (i_{\psi}\sigma_1)$ for some $\psi\in T$, we have
    $i_XA_1(\sigma_2)=-i_{\psi} i_X\sigma_2$ and so $A_1(\sigma_2)=i_{\psi} \sigma_2$. Finally, taking $\sigma_2$ decomposable we obtain
    \[0=A(\sigma_2)\sigma_2=A_0(\sigma_2)\sigma_2-(i_{e}i_{\psi}\sigma_2)\sigma_2,\]
    from which we get $A_0(\sigma_2)= i_{e}i_{\psi}\sigma_2$ for an arbitrary 2-form $\sigma_2$. Putting things together,
    \[A(\sigma_2)=i_{e}i_{\psi} \sigma_2+i_{\psi}\sigma_2\otimes e-(F_1\wedge\sigma_2-F_2\wedge i_{e}\sigma_2)-F_4\wedge \sigma_2.\]
    
    \subsection*{5-forms}
    The above analysis yields \[-B(\sigma_2\wedge\sigma_2')=i_{e}i_{\psi} (\sigma_2\wedge\sigma_2')-2F_1\wedge\sigma_2\wedge\sigma_2'+F_2\wedge i_{e}(\sigma_2\wedge\sigma_2'),\]
    and so $B(\sigma_4)=i_{\psi} i_{e}\sigma_4+2F_1\wedge\sigma_4-F_2\wedge i_{e}\sigma_4$. Using this result, we can take $u=X$, $v=\sigma_5$ to get
    \[i_{\psi} i_{e} i_X\sigma_5+2F_1\wedge i_X\sigma_5-F_2\wedge i_{e} i_X\sigma_5=2i_XF_1\wedge\sigma_5-i_X F_2\wedge i_{e}\sigma_5+i_X(A_3(\sigma_5)+A_6(\sigma_5))+e(i_X A_1(\sigma_5)),\]
    implying $A_1(\sigma_5)=0$, $A_3(\sigma_5)=i_{\psi} i_{e} \sigma_5$, and $A_6(\sigma_5)=-2F_1\wedge \sigma_5+F_2\wedge i_{e}\sigma_5$. Finally, for any $\sigma_5$ there exists $\sigma_2\neq 0$ such that $\sigma_2\bar\otimes\sigma_5=0$. For these forms we have $0=A(\sigma_2) \sigma_5+A(\sigma_5)\sigma_2$. Taking the 2-form part of this expression, we have $0=A_0(\sigma_5)\sigma_2$ and hence $A_0(\sigma_5)=0$ for any $\sigma_5$. Thus,
    \[A(\sigma_5)=i_{\psi} i_{e} \sigma_5-(2F_1\wedge \sigma_5-F_2\wedge i_{e}\sigma_5).\]

  \section{Explicit form of the twisted bracket}\label{ap:fullbkt}
    The full bracket \eqref{eq:fulltwisted} reads
    \begin{align*}
    [x+\tau_1&+\tau_0+\tau_2+\tau_4+\tau_5,\;x'+\tau'_1+\tau'_0+\tau'_2+\tau'_4+\tau'_5]=\mc L_x x'\\
    &+(\mc L_x \tau_1'-i_{x'}d\tau_1+\beta_0 \tau_0\tau_1'+\beta_0i_{x'}\tau_2+i_{x'}i_x\beta_3+\tau_1'i_x\alpha_1+\tau_1'i_x\beta_1-i_{x'}(\alpha_1\wedge\tau_1)-i_{x'}(\beta_1\wedge\tau_1))\\
    &+(\mc L_x \tau_0'-\mc L_{x'}\tau_0+\varphi \,i_{x'}\tau_1+i_{x'}i_x\alpha_2-\tau_0'i_x\beta_1+\tau_0i_{x'}\beta_1)\\
    &+(\mc L_x \tau_2'-i_{x'}d\tau_2+d\tau_0\wedge \tau_1'+\tau_0'd\tau_1+\beta_0\tau_0\tau_2'-\beta_0\tau_0'\tau_2-\varphi\,\tau_1\wedge \tau_1'+i_{x'}i_x\alpha_4-\tau_0'i_x\beta_3\\
    &+\tau_1'\wedge i_x\alpha_2+\tau_2'i_x\alpha_1+\tau_0i_{x'}\beta_3-\tau_0\beta_1\tau_1'+i_{x'}(\alpha_2\wedge\tau_1)+\alpha_1\tau_1\tau_0'+\beta_1\tau_1\tau_0'-i_{x'}(\alpha_1\wedge\tau_2))\\
    &+(\mc L_x \tau_4'-i_{x'}d\tau_4+d\tau_1\wedge\tau_2'+d\tau_2\wedge \tau_1'+2\beta_0 \tau_0\tau_4'-\beta_0\tau_2\wedge\tau_2'+2\beta_0i_{x'}\tau_5+\tau_1'\wedge i_x\alpha_4\\
    &-\tau_2'i_x\beta_3+2\tau'_4i_x \alpha_1+\tau'_4i_x\beta_1-\tau_0\beta_3\tau_1'+i_{x'}(\alpha_4\wedge\tau_1)-\alpha_2\wedge\tau_1\wedge\tau_1'+\alpha_1\wedge\tau_1\wedge\tau_2'\\
    &+\beta_1\wedge\tau_1\wedge\tau_2'+i_{x'}(\beta_3\wedge\tau_2)+\alpha_1\wedge\tau_2\wedge\tau_1'-2i_{x'}(\alpha_1\wedge\tau_4)-i_{x'}(\beta_1\wedge\tau_4))\\
    &+(\mc L_x \tau_5'+d\tau_0\wedge\tau_4'+\tau_0'd\tau_4-d\tau_2\wedge\tau_2'+2\beta_0\tau_0\tau_5'-2\beta_0\tau_0'\tau_5-\varphi\,\tau_1\wedge \tau_4'+\tau_2'\wedge i_x\alpha_4\\
    &-\tau_4'\wedge i_x\alpha_2+2\tau_5'i_x\alpha_1+\tau_0\beta_3\wedge\tau_2'-\tau_0\beta_1\wedge\tau_4'-\tau_0'\alpha_4\wedge \tau_1+\alpha_2\wedge\tau_1\wedge\tau_2'-\tau_0'\beta_3\wedge\tau_2\\
    &-\alpha_1\wedge\tau_2\wedge\tau_2'+2\tau_0'\alpha_1\wedge\tau_4+\tau_0'\beta_1\wedge\tau_4),
  \end{align*}
  where the twists
  \[\varphi\in C^\infty(M),\;\beta_0\in C^\infty(M),\;\alpha_1\in \Omega^1(M),\;\beta_1\in\Omega^1(M),\;\alpha_2\in\Omega^2(M),\;\beta_3\in\Omega^3(M),\;\alpha_4\in\Omega^4(M),\]
  satisfy the constraints
  \[\varphi\beta_0=0 \qquad d\varphi-2\varphi\beta_1-\varphi\alpha_1=0\qquad d\beta_0+\beta_1\beta_0=0\qquad d\alpha_1-\alpha_2\beta_0=0\qquad d\beta_1=0\]
  \[d\alpha_2-\alpha_2\wedge\beta_1-\varphi\beta_3=0\qquad d\beta_3-\beta_3 \wedge\alpha_1-\alpha_4\beta_0+\beta_1\wedge\beta_3=0\qquad d\alpha_4+\alpha_4\wedge\alpha_1-\alpha_2\wedge\beta_3=0.\]
}

\end{document}